\documentclass[12pt,a4paper]{article}

\usepackage[english]{babel}
\usepackage{amsmath,amsthm,amssymb}
\usepackage{bbm,graphicx,psfrag,cite}

\numberwithin{equation}{section}

\newcommand{\Id}{\mathbbm{1}}

\newcommand{\Z}{\mathbbm{Z}}

\newcommand{\R}{\mathbbm{R}}
\newcommand{\Pb}{\mathbbm{P}}

\newcommand{\dx}{\mathrm{d}}

\newcommand{\Pf}{\mathrm{Pf}}
\newcommand{\Var}{\mathrm{Var}}
\newcommand{\Tr}{\mathrm{Tr}}

\newcommand{\Ai}{\mathrm{Ai}}

\newcommand{\I}{{\rm i}}

\numberwithin{equation}{section}

\title{The Airy$_1$ process is not the limit of the largest eigenvalue
  in GOE matrix diffusion}
\author{Folkmar Bornemann\thanks{Technische Universit\"at M\"unchen,
    e-mail: bornemann@ma.tum.de},
Patrik L. Ferrari\thanks{Weierstrass Institute, WIAS Berlin, e-mail: ferrari@wias-berlin.de},
Michael Pr\"ahofer\thanks{Technische Universit\"at M\"unchen, e-mail: praehofer@ma.tum.de}}

\date{June 20, 2008}

\begin{document}
\maketitle \sloppy

\begin{abstract}
Using a
systematic approach to evaluate Fredholm determinants
numerically, we provide convincing evidence that the Airy$_1$-process,
arising as a limit law in stochastic surface growth, is not the
limit law for the evolution of the largest
eigenvalue in GOE matrix diffusion.
\end{abstract}

\section{Introduction}

One of the unsolved problems in random matrix theory is to understand
the law for the largest eigenvalue in GOE matrix diffusion. Let $M(t)$
be a matrix-valued stationary process on real symmetric matrices of
size $N\times 
N$ satisfying: (i) the one-time
  distribution of $M(t)$ is given by the Gaussian Orthogonal Ensemble
  (GOE),
(ii) the (independent) entries of $M(t)$ are independent
  stationary Ornstein-Uhlenbeck processes in time.
The corresponding process for the ordered eigenvalues
$\lambda_{N,j}(t)$, $j=1,\dots,N$ is Dyson's Brownian motion with
$\beta=1$. The stationary distribution of the largest eigenvalue,
$\lambda_{N,N}(t)$, can be expressed fairly explicit, and its limiting
distribution, under proper rescaling as $N\to\infty$, is the GOE Tracy-Widom
distribution \cite{TW96}. However, no simple expression for the joint
distribution of the largest eigenvalue at two different times is
known.  To be specific, one can ask for the covariance of the largest
eigenvalue ${\rm Cov}\big(\lambda_{N,N}(t),\lambda_{N,N}(0)\big)$ and
its asymptotic behavior as $N\to\infty$.

For the related case of GUE matrix diffusion, i.e., when $M(t)$ is a
hermitian matrix and the stationary distribution is given by the Gaussian
Unitary Ensemble (GUE), the corresponding question is answered. In
this case the law 
for the eigenvalues is given by Dyson's Brownian motion with
$\beta=2$ and the limiting process of the properly rescaled largest
eigenvalue is the Airy process. This process first arose in the context of
one-dimensional stochastic surface growth with curved macroscopic
shape~\cite{PS02} for the so-called polynuclear growth (PNG) model.

This raised the question, whether the limit process of the largest
eigenvalue in GOE matrix diffusion can also be obtained from a growth
process. A strong candidate was  one-dimensional
growth starting from a flat substrate, since in this case the limiting
one-point distribution is the same as for GOE matrix
diffusion~\cite{PS00}.   This
correspondence was partially extended to a multilayer version of flat
growth with non-intersecting height lines. It was shown in~\cite{Fer04}
that the point process of the multilayer at a fixed position and the
point process of the GOE ensemble at the edge of the spectrum have the
same asymptotic law.
 
The analogue of the Airy process for flat growth was discovered by Sasamoto 
in a growth model related to PNG \cite{Sas05}. 
Since its defining kernel at equal times is, in a certain sense, the
square root of the standard Airy kernel \cite{FS05b}, it was baptized
the ``Airy$_1$ process''. Accordingly, for better distinction, we call
the standard Airy process ``Airy$_2$ process'' in the rest of the paper.

In \cite{BFPS06} two conjectures have been formulated. The first predicted
that the Airy$_1$ process is also the limit process for the PNG model
with flat initial conditions, which subsequently has been proven in
\cite{BFS07}. The second claimed that the Airy$_1$ process is also the
limit of the largest eigenvalue in GOE matrix diffusion ($\beta=1$
Dyson's Brownian motion).

In this paper we show that the second conjecture does not hold. To
this end we compare the two-point 
functions of the Airy$_1$ process and of the largest eigenvalue in GOE
matrix diffusion for different matrix sizes. 

The joint distribution functions for the Airy processes are given in
terms of Fredholm determinants of integral operators. To evaluate
these Fredholm determinants we employ a 
numerical scheme, recently developed by one of the authors
\cite{Born08}, which in itself is of general interest.
For matrix diffusion we use straightforward Monte-Carlo
simulations on large matrices. 

The comparison shows that the correlation function for GOE matrix
diffusion differs, in the limit of large matrices, from the one for
Airy$_1$. In contrast, in the case of GUE matrix
diffusion, the corresponding numerical calculations perfectly
illustrate the known convergence to the Airy$_2$ process.

\section{Polynuclear growth model}
In this section we present the polynuclear growth model in $1+1$
dimensions and give known results relevant for the discussion. We
refer to the original papers for more details.

\subsubsection*{The model and its multilayer extension}
We briefly define the polynuclear growth (PNG) model on a
one-dimensional substrate. At time $t$, the surface is described by an
integer-valued height function $x\mapsto h(x,t)\in \Z$, $x\in \R,t\in
\R_+$, with steps of size $1$, which is taken to be upper
semicontinuous, i.e.,
$\lim\limits_{x\to x_0}h(x,t)\leq h(x_0,t)$ for all $x_0,t$. Thus
the surface consists of up-steps
($\lrcorner\hspace{-0.15em}\ulcorner$) and down-steps
($\urcorner\hspace{-0.15em}\llcorner$). The dynamics of these steps has a
deterministic and a stochastic part:
\begin{itemize}
\item[(i)] up- (down-) steps move to the left (right) with unit
  speed. When a down-step and an up-step collide they simply disappear.

\item[(ii)] pairs of up- and down- steps at the same point (spikes)
  are produced by random nucleation events with some given intensity.
  The up- and down-steps of the spikes then spread out with unit
  speed according to (i).
\end{itemize}

The multilayer extension of the PNG model~\cite{PS02} is the
following. Instead of a single height function $h(x,t)$ we have a set
of height functions $\{h_\ell(x,t),\ell\leq 0\}$, with the initial
condition  $h_\ell(x,0)=\ell$, for all $x\in\R$. The dynamics of
$h_0(x,t)$ is the same as for the original $h(x,t)$. For the remaining
lines (i) applies as for $h_0(x,t)$. Rule (ii) is modified insofar,
that for $h_\ell(x,t)$, $\ell\leq-1$, nucleation events are not
produced at random, but whenever there is a collision of a pair of
steps in level $\ell+1$ at $(x,t)$, a spike is produced in level
$\ell$ at $(x,t)$.
By construction the lines
do not intersect and one associates an (extended) point process
$\eta$ on $\R\times\Z$, by
\begin{equation}\label{eqPP}
\eta(x,j)=\left\{\begin{array}{ll} 1,& h_\ell(x,t)=j\textrm{ for some
}\ell\leq 0,\\ 0,&\textrm{otherwise}.
\end{array} \right.
\end{equation}

\subsubsection*{The PNG droplet}
Consider a flat initial substrate $h(x,0)=0$, $x\in\R$. The PNG droplet
is obtained when the nucleations form a Poisson point process in
space-time with intensity $\rho(x,t)=2$ for $|x|\leq t$ and
$\rho(x,t)=0$ otherwise. For large growth time $t$, the interface has
the shape of a droplet, namely the deterministic limit,
\begin{equation}
h_{\rm ma}(\xi):=\lim_{t\to\infty}t^{-1}h(\xi
t,t)=2\sqrt{1-\xi^2}\Id_{(|\xi|\leq 1)}.
\end{equation}
The fluctuations of the height function grow as $t^{1/3}$ and the
correlation length as $t^{2/3}$. Therefore, the edge scaling of the
(multilayer) height functions around the origin, $x=0$, is given by
\begin{equation}\label{eqEdgeDroplet}
h^{\rm droplet}_\ell(u,t):=\frac{h_\ell(u
t^{2/3})-th_{\rm ma}({u}{t^{-1/3}})
}{t^{1/3}}.
\end{equation}
For the PNG droplet, the
point process associated to the multilayer is determinantal. Moreover,
rescaled as in (\ref{eqEdgeDroplet}), it converges in the large $t$
limit to the Airy field~\cite{PS02}, defined by the n-point
correlation functions
\begin{equation}
\rho^{(n)}(u_1,s_1;\ldots;u_n,s_n)=\det(K_{{\cal
A}_2}(u_i,s_i;u_j,s_j))_{1\leq i,j\leq n},
\end{equation}
where
\begin{equation}
K_{{\cal A}_2}(u,s;u',s') = \left\{\begin{array}{ll} \displaystyle
\phantom{+}\int_0^\infty\dx \lambda
e^{(u'-u)\lambda}\Ai(s+\lambda)\Ai(s'+\lambda), & u'\leq u, \\[1em]
\displaystyle -\int_{-\infty}^0\dx \lambda
e^{(u'-u)\lambda}\Ai(s+\lambda)\Ai(s'+\lambda), & u'>u.
\end{array}
\right.
\end{equation}

Denote by ${\cal A}_2(u)$ the highest point of the Airy field at
position $u$. It can be seen as a process $u\mapsto {\cal A}_2(u)$ and
it is called the Airy$_2$ process. The convergence of the extended
point process to the Airy field implies in particular that~\cite{PS02}
\begin{equation}
\lim_{t\to\infty}h^{\rm droplet}_0(u,t)={\cal A}_2(u).
\end{equation}
The joint distributions of the Airy$_2$ process are given by
Fredholm determinants: for any given $u_1<u_2<\ldots<u_m$, and
$s_1,\ldots,s_m\in\R$,
\begin{equation}\label{A2FredDet}
\Pb\left(\bigcap_{k=1}^m \{{\cal A}_2(u_k)\leq
s_k\}\right)=\det(\Id-\chi_s K_{{\cal A}_2}
\chi_s)_{L^2(\{u_1,\ldots,u_m\}\times \R)},
\end{equation}
where $\chi_s(u_k,x)=\Id_{(x> s_k)}$.  This expression allows to
determine some properties of the covariance
\begin{equation}\label{CovDefAiry2}
g_2(u):={\rm Cov}({\cal A}_2(u),{\cal A}_2(0)),
\end{equation}
namely
\begin{equation}\label{CovPropertiesAiry2}
g_2(0)=\Var({\cal A}_2(0))= 0.81320\ldots,\quad g_2'(0)=-1,
\end{equation}
and the asymptotics for large $u$ \cite{AvM03,Wid03},
\begin{equation}
  \label{eq:CovAsymptAiry2}
  g_2(u)=\frac1{u^2}+\frac{c}{u^4}+{\cal O}(u^{-6}),
\end{equation}
with the constant $c=-3.542\dots$, evaluated numerically from an
explicit expression in terms of the Hastings-McLeod solution of
Painlev\'e II \cite{Born08}.
\subsubsection*{The flat PNG}
Consider a flat initial substrate $h(x,0)=0$, $x\in\R$, and run the PNG
dynamics with constant nucleation intensity, say $\rho(x,t)=2$ for all
$x\in\R$, $t\geq 0$. Then the limit shape is flat, $h_{\rm
ma}(\xi)=2$. Thus the edge scaling is
\begin{equation}\label{eqEdgeFlat}
h^{\rm flat}_\ell(u,t):=\frac{h_\ell(u t^{2/3})-2t}{t^{1/3}}.
\end{equation}

For the flat PNG, the correlation structure of the multilayer version
is not known, but a few results are available.

(a) In the large time limit the point process corresponding to
(\ref{eqEdgeFlat}) for a fixed value of $u$ converges to a Pfaffian
point process~\cite{Fer04}
whose $n$-point correlation functions are given by
\begin{equation}\label{PfaffianPNG}
\rho^{(n)}(s_1,\ldots,s_n)=2^{2n/3}\Pf(G^{\rm
GOE}(2^{2/3}s_i;2^{2/3}s_j))_{1\leq i,j\leq n}.
\end{equation}
$G^{\rm GOE}$ is a $2\times 2$ matrix kernel (for an explicit
expression see, e.g.\ (2.9) in~\cite{Fer04}) and $\Pf$ is the Pfaffian
($\Pf(A)=\sqrt{\det(A)}$ for an antisymmetric matrix $A$). This kernel
also occurs for GOE random matrices in the large matrix limit at the
edge of the spectrum.

(b) Recently, it has been proven that the Airy$_1$ process describes
the limit of the top line of the multilayer flat PNG~\cite{BFS07b},
namely
\begin{equation}
\lim_{t\to\infty} h^{\rm flat}_0(u,t)=2^{1/3} {\cal A}_1(u/2^{2/3}).
\end{equation}

The joint distributions of the Airy$_1$ process are given by 
Fredholm determinants: for any given $u_1<u_2<\ldots<u_m$, and
$s_1,\ldots,s_m\in\R$,
\begin{equation}\label{A1FredDet}
\Pb\left(\bigcap_{k=1}^m \{{\cal A}_1(u_k)\leq
s_k\}\right)=\det(\Id-\chi_s K_{{\cal A}_1}
\chi_s)_{L^2(\{u_1,\ldots,u_m\}\times \R)},
\end{equation}
where $\chi_s(u_k,x)=\Id(x> s_k)$, and the kernel $K_{{\cal A}_1}$ is
defined by
\begin{align}
K_{{\cal A}_1}(u,s;u',s')&=
\Ai(s+s'+(u-u')^2)\exp\left((u'-u)(s+s')+\tfrac23 (u'-u)^3\right)
\nonumber \\
&-\frac{1}{\sqrt{4\pi(u'-u)}}\exp\left(-\frac{(s'-s)^2}{4(u'-u)}\right)\Id(u'>u)
\end{align}

Some properties of the Airy$_1$ process like the short
and long time behavior of the covariance are known. We refer to the
review~\cite{Fer07} for details. In particular, the short-time
behavior of the covariance of the Airy$_1$ process,
\begin{equation}\label{CovDefAiry1}
g_1(u):={\rm Cov}({\cal A}_1(u),{\cal A}_1(0)),
\end{equation}
satisfies
\begin{equation}\label{CovProperties}
g_1(0)=\Var({\cal A}_1(0))\simeq 0.402\ldots,\quad g_1'(0)=-1.
\end{equation}

\section{Dyson's Brownian motion}
Dyson's Brownian motion~\cite{Dys62} describes the diffusion of $N$
 mutually repelling particles with positions $\lambda_j(t)$,
 $j=1,\dots,N$, at time $t$ on the real line in a harmonic potential,
\begin{eqnarray}\label{eq:DBMsde}
\dx\lambda_j(t)=\left(-\gamma\lambda_j(t) +\frac\beta2
\sum\limits_{i\neq j}\frac{1}{\lambda_j(t)-\lambda_i(t)}\right)\dx t+ \dx
b_j(t),\quad j=1,\ldots,N,
\end{eqnarray}
the $b_j(t)$ being independent standard Brownian motions with
$\Var(b_j(t))=t$.
Let $P(\lambda)$ denote the probability distribution
of particle positions $\lambda=(\lambda_1,\ldots,\lambda_N)$. It
satisfies the diffusion equation
\begin{eqnarray}\label{eq:DBMpde}
\frac{\partial}{\partial t}P(\lambda)= \sum\limits_{j=1}^N
\frac{\partial}{\partial\lambda_j} \left(\gamma\lambda_j
P(\lambda)-\frac{\beta}{2} \sum\limits_{i\neq
j}\frac{1}{\lambda_j-\lambda_i}P(\lambda)\right) +\frac12
\frac{\partial^2}{\partial \lambda_j^2}P(\lambda).
\end{eqnarray}
The stationary distribution is given by
\begin{eqnarray}\label{eq:DBMstationary}
P(\lambda)=\frac1Z|\Delta(\lambda)|^\beta
\exp\bigg(\!\! -\gamma\sum_{j=1}^N\lambda_j^2\bigg),
\end{eqnarray}
with $\Delta(\lambda)=\prod\limits_{1\leq i<j \leq
N}(\lambda_j-\lambda_i)$, and $Z$ the normalization.

In his original work, Dyson linked the special values $\beta=1,2,4$ to
the eigenvalue GOE, GUE and GSE random matrices, respectively. In
these cases, $\lambda_j(t)$ is the $j$-th smallest eigenvalue of a
random matrix $M(t)$ diffusing according to an Ornstein-Uhlenbeck
process on real symmetric, hermitian or symplectic matrices,
respectively.

We describe the correspondence only in the cases $\beta=1$ (GOE) and
$\beta=2$ (GUE). Let $b^\alpha_{ij}(t)$, $1\leq i,j\leq N$, $\alpha=1,2$, be
independent standard Brownian motions. In the GOE case one sets
$b_{ij}(t)=b^1_{ij}(t)\in\R$, in the GUE case one sets
$b_{ij}(t)=b^1_{ij}(t)+\I b^2_{ij}(t)\in\mathbb{C}$.
Let
$B_{ij}(t)=\frac12\big(b_{ij}(t)+\overline{b_{ji}(t)}\big)$. Now $B(t)$
is a Brownian motion on the space of real symmetric, resp., hermitian
matrices, which is invariant with respect to orthogonal, resp. unitary
rotations. For GOE the independent entries
are $B_{ij},j\geq i$, while for GUE the independent
entries are $B_{ii}$ and
$\mathrm{Re}(B_{ij}),\mathrm{Im}(B_{ij})$, $j>i$. These real-valued
independent entries perform Brownian motions, with variance $t$ on the
diagonal 
and variance $t/2$ for the remaining entries. Now let $M(t)$ be
the stationary Ornstein-Uhlenbeck process defined by 
\begin{eqnarray}
  \label{eq:RMstationary}
  \dx M(t)=-\gamma M(t)\dx t+\dx B(t).
\end{eqnarray}
The stationary distribution is proportional to $\exp(-\gamma\Tr
(M^2))$ in both cases, GOE and GUE. 
By integrating over the angular variables, one gets the stochastic
evolution of eigenvalues as in 
(\ref{eq:DBMstationary}).

Let us mention here, that the parameter $\gamma$ is in fact
irrelevant. Multiplying the eigenvalues by $\sqrt{\gamma}$ and
rescaling time by $\gamma^{-1}$ one can always arrange for
$\gamma=1$. We kept this parameter throughout the formulas to
facilitate comparisons with the literature, where different, sometimes
$N$-dependent, conventions for $\gamma$ have been adopted. The most 
common choice, $\gamma=1$, leads to the standard Hermite kernel with
Hermite polynomials orthogonal with respect to the weight $e^{-x^2}$. 

\subsubsection*{GUE diffusion and Airy process}

In the case $\beta=2$, the point process associated to the
ordered eigenvalues, $\lambda_j(t)$ of $M(t)$
is determinantal, defined by the extended Hermite kernel \cite{Jo05b}. The edge scaling at
the upper edge of the spectrum is given by
\begin{eqnarray}\label{eq:rescaledGUE}
\lambda_{N,j}^{\rm GUE}(u)=\sqrt{2\gamma}N^{1/6}
\left(\lambda_j(u/(\gamma N^{1/3}))-\sqrt{2N/\gamma}\right).
\end{eqnarray}
Under this rescaling, the kernel of the
corresponding point process converges to the Airy kernel $K_{{\cal
A}_2}$ as $N\to\infty$~\cite{AvM03}. This allows to show the 
convergence of the rescaled largest eigenvalue, $\lambda_{N,N}^{\rm
  GUE}(u)$, to the Airy$_2$ 
process,
\begin{equation}\label{eq:ConvToAiry}
\lim_{N\to\infty}\lambda_{N,N}^{\rm GUE}(u)={\cal A}_2(u),
\end{equation}
in the sense of convergence of finite-dimensional distributions \cite{Jo03b}.
The finite-$N$ covariance of the largest eigenvalue is denoted by $f_N^{\rm GUE}$,
\begin{equation}\label{CovDefGUE}
f_N^{\rm GUE}(u)={\rm Cov}\left(\lambda_{N,N}^{\rm GUE}(u),\lambda_{N,N}^{\rm GUE}(0)\right).
\end{equation}
Obviously one expects that $\lim_{N\to\infty}f_N^{\rm
  GUE}(u)=g_2(u)$. To prove this rigorously, convergence of moments is
  needed in   (\ref{eq:ConvToAiry}), a result which is currently not
  available. 

\subsubsection*{GOE diffusion}

For $\beta=1$, the GOE case, explicit expressions for dynamical
correlations are not known. Nevertheless one expects an analogous
behavior as for GUE diffusion. In the edge scaling of the ordered eigenvalues
$\lambda_j(t)$ of $M(t)$ in the GOE case, 
one has two free parameters (time and space scaling). In order to
check the hypothesis that the Airy$_1$ process describes the evolution
of the largest eigenvalue for the GOE case, we choose the free
parameters by matching the covariance and its derivative at zero, see
(\ref{CovProperties}). We obtain
\begin{eqnarray}\label{eq:rescaledGOE}
\lambda_{N,j}^{\rm GOE}(u)=\sqrt{\gamma}N^{1/6}\left(\lambda_j(2u/(\gamma N^{1/3}))-\sqrt{N/\gamma}\right).
\end{eqnarray}
As anticipated while speaking of the flat PNG, the point process of the edge-rescaled GOE eigenvalues at a fixed time (the one associated to $\{\lambda_{N,j}^{\rm GOE}(0),1\leq j\leq N\}$) converges to a Pfaffian point process with $n$-point correlation given by
\begin{equation}\label{PfaffianGOE}
\rho^{(n)}(s_1,\ldots,s_n)=2^n\Pf(G^{\rm GOE}(2s_i,2s_j))_{1\leq i,j\leq n}
\end{equation}
in the $N\to\infty$ limit. We denote by $f_N^{\rm GOE}$ the finite-$N$ covariance of the largest eigenvalue,
\begin{equation}\label{CovDefGOE}
f_N^{\rm GOE}(u)={\rm Cov}\big(\lambda_{N,N}^{\rm
  GOE}(u),\lambda_{N,N}^{\rm GOE}(0)\big). 
\end{equation}
The scaling in (\ref{eq:rescaledGOE}) is chosen such that 
${f_N^{\rm GOE}}'(0)=-1$ and $f_N^{\rm  GOE}(0)\to g_1(0)$ as $N\to\infty$.
As in the GUE case one expects the limit $f_\infty^{\rm
  GOE}(u)=\lim_{N\to\infty}f_N^{\rm  GOE}(u)$ to exist. In the next
section we address the question whether  $f_\infty^{\rm GOE}(u)$ equals
$g_1(u)$. 

\section{Numerical results} \label{sec:numresults}

The Airy$_1$ process was regarded as a candidate for the limit of the
rescaled largest GOE eigenvalue process \cite{Sas05,BFPS06}, because
of the known properties of these two processes. Both are stationary
processes with the same one-point distribution, and the conjectured short
time behaviors coincide, too.  Furthermore, as explained above, the
underlying multiline ensembles have the same limiting single time
distribution as point processes on $\R$ (see (\ref{PfaffianPNG}) and
(\ref{PfaffianGOE})). The final ingredient for the guess is that the
connections carries over the multilines picture in the $\beta=2$ case.

In lack of more analytical input, we looked for an answer to
the question, whether the Airy$_1$ process is the limit of the $\beta=1$
Dyson's 
Brownian motion by numerical means. 
We decided to compare the large $N$ limit of (\ref{CovDefGOE}) with
the covariance of the Airy$_1$ process (\ref{CovDefAiry1}). The
quantities in question are 
not straightforwardly accessible. For the 
Airy$_1$ process one needs to evaluate Fredholm determinants. For
Dyson's Brownian motion we performed
 Monte-Carlo simulations on the eigenvalues of coupled GOE
matrices directly.

\subsubsection*{Covariances of the Airy$_1$ and Airy$_2$ processes}
The key point of the numerical computation is the evaluation of the
Fredholm determinants of the two-point joint distributions for the
Airy$_1$ and Airy$_2$ processes,
eqs. (\ref{A2FredDet}),~(\ref{A1FredDet}). This is explained in
details in the  
recent paper~\cite{Born08}. Let us briefly describe the procedure.

Basic ingredient is a Nystr\"om-type approximation of integral operators
by an $n$-point quadrature formula for integrals over the interval
$(s,\infty)$ that gives exponential convergence rates for holomorphic
integrands. Such a formula can be based on the holomorphic transformation
\[
\phi_s : (0,1) \to (s,\infty),\quad \xi \mapsto s + 10 \tan(\pi\xi/2),
\]
followed by Gauss--Legendre quadrature on the interval $(0,1)$ with
weights $w_j$ and points $\xi_j$ $(j=1,\ldots,n)$. This way we obtain
\[
\int_{s_k}^\infty f(x) \approx \sum_{j=1}^n w_j\phi'_{s_k}(\xi_j) \,
f (\phi_{s_k}(\xi_j)) = \sum_{j=1}^n w_{kj} f(x_{kj}).
\]
The Fredholm determinants (\ref{A2FredDet}) and (\ref{A1FredDet}) are
now approximated by the $mn\times mn$-dimensional determinant
\begin{equation}\label{FredDetApprox}
\det(\Id-\chi_s K_{{\cal A}_\mu}
\chi_s)_{L^2(\{u_1,\ldots,u_m\}\times \R)} \approx \det
\begin{pmatrix}
\Id - A_{11} & A_{12} & \cdots & A_{1m} \\*[1mm]
A_{21} & \Id - A_{22} & \cdots & A_{2m} \\*[1mm]
\vdots & \vdots & &  \vdots \\*[1mm]
A_{m1} & A_{m2} & \cdots & \Id - A_{mm}
\end{pmatrix}
\end{equation}
where the submatrices $A_{ij} \in \R^{n\times n}$ ($i,j=1,\ldots,m$) are
defined by
\begin{equation}
(A_{ij})_{pq} = w_{ip}^{1/2} K_{{\cal A}_\mu}(u_i,x_{ip};u_j,x_{jq})
w_{jq}^{1/2}\qquad (p,q=1,\ldots,n).
\end{equation}
Theorem~8.1 of \cite{Born08} shows that the approximation error in
(\ref{FredDetApprox}) decays exponentially with $n$, that is, like
$O(\rho^{-n})$ for some constant $\rho>1$. Thus, doubling $n$ doubles
the number of correct digits; a fact on which simple strategies for
adaptive error control can be based \cite{Born04}. Additionally, the
level of round-off error can be controlled as described in \cite{Born08}.
It turns out that the two-point ($m=2$) joint distribution can be
calculated to an absolute precision of $10^{-14}$ using $n$ quadrature
points with $n$ between $20$ and $100$, depending on the specific
values of $u$, $s_1$, and $s_2$. The CPU time for a single evaluation of
the joint distribution is well below a second (using a 2~GHz PC).

The covariances $g_1(u)$ and $g_2(u)$ were calculated by first truncating
the integrals, then integrating by parts (to avoid numerical differentiation),
and finally using Clenshaw--Curtis quadrature. A truncation at $\pm 10$ (except
for  small $u\leq 0.05$ in the case of $g_1$, where larger integration
intervals are necessary) and 100 quadrature points in each of the two dimensions
are sufficient to secure an absolute precision of $10^{-10}$.

The code for
calculating the two-point joint distributions and
the covariances $g_1$ and $g_2$ can be obtained from the first
author upon request.

\subsubsection*{Monte Carlo Simulation of Random Matrices}
To get an estimate for the covariance of the largest eigenvalue of GUE
and GOE matrices, we performed straightforward Monte-Carlo
simulations.
A collection of matrices $C_k$, $k=0,\dots,K$, independently
distributed according to the stationary distribution of
(\ref{eq:RMstationary}) can be easily produced with standard pseudo
random generators, since each $C_k$ consists of independently normally
distributed entries. Fixing a time step $\Delta t$, it is easy to see
that for the  stationary process $M(t)$
governed by (\ref{eq:RMstationary}) the joint distribution of
$\left(M(k\Delta t)\right)_{0\leq k\leq K}$ is the same as for the matrices
$M_k$, defined by 
\begin{eqnarray}
M_0=C_0,
\qquad M_{k}=e^{-\gamma \Delta t}M_{k-1}+\sqrt{1-e^{-2\gamma \Delta
    t}}C_{k},\quad 1\leq k\leq K.
\end{eqnarray}   
Now, one numerically determines the
largest eigenvalues of the 
$M_k$ and rescales according to eqs. (\ref{eq:rescaledGUE}), resp.,
(\ref{eq:rescaledGOE}). This yields realizations of $\lambda_{N,N}^{\rm
  GOE}(u)$ and $\lambda_{N,N}^{\rm
  GUE}(u)$ at equidistant times $u$. The
empirical auto-covariance of these time series
gives an estimate for the covariance functions $f_N^{\rm GOE}(u)$ and $f_N^{\rm
  GUE}(u)$ at discrete values of $u$.
The data presented here are for $N=64$, $128$, and $256$ with
$\gamma=1/2$ and $\Delta
t=\frac12N^{-1/3}$. We chose $K=10^6$, and produced up to $100$
independent realizations of the time series in each case. This allows to
obtain an error estimate for each data point.

\subsubsection*{Discussion}
\label{discussion}

\begin{figure}
\begin{center}
\includegraphics[height=8cm]{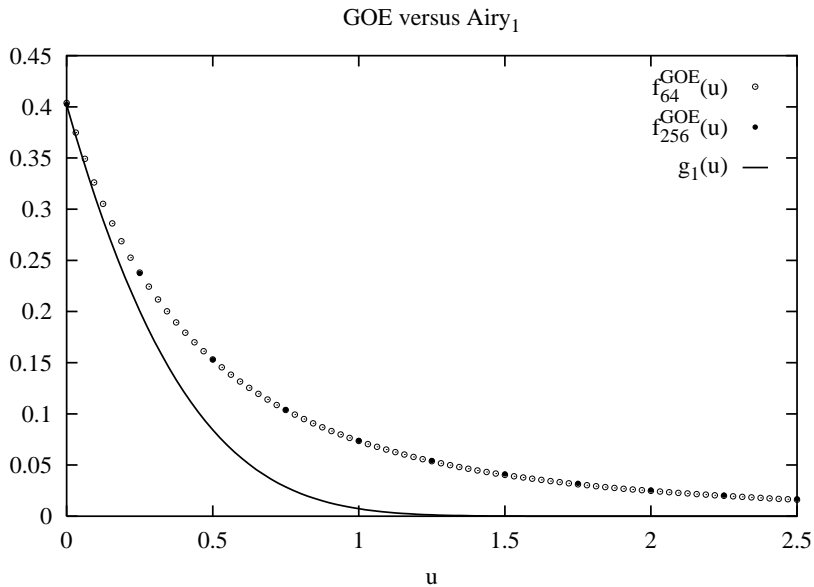}
\caption{The covariance $g_1$ of the Airy$_1$ process (line) versus the
  one of the largest eigenvalue for GOE matrix diffusion, $f_N^{\rm
    GOE}$, for $N=64,256$.}
\label{FigCovarianceGOE}
\end{center}
\end{figure}

In Figure~\ref{FigCovarianceGOE} we compare the covariance
(\ref{CovDefAiry1}) of the Airy$_1$ process and the one of the largest
eigenvalue for GOE matrix diffusion (\ref{CovDefGOE}) for $N=64$ and
$N=256$. One clearly sees that they do not agree.

\begin{figure}
\begin{center}
\includegraphics[height=8cm]{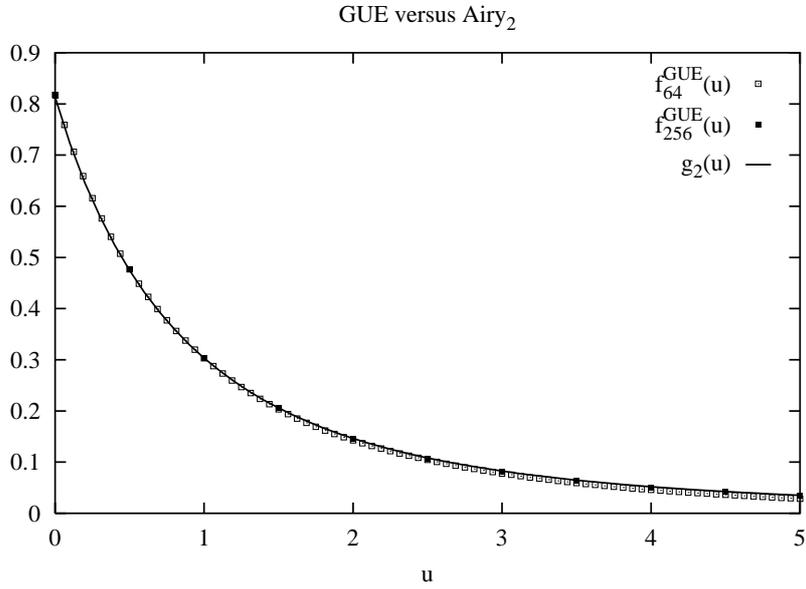}
\caption{The covariance $g_2$ of the Airy$_2$ process (line) versus the
  one of the largest eigenvalue for GUE matrix diffusion, $f_N^{\rm
    GUE}$, for $N=64,256$.}
\label{FigCovarianceGUE}
\end{center}
\end{figure}

Increasing the matrix dimension does not change sensibly the result;
namely, the results for $N=128$ agree to plotting accuracy with the
one for $N=256$
in Figure~\ref{FigCovarianceGOE} and therefore are not shown. In
comparison, in Figure~\ref{FigCovarianceGUE} we plot the covariance 
(\ref{CovDefAiry2}) of the Airy$_2$ process and the one of the largest
eigenvalue for GUE matrix diffusion (\ref{CovDefGUE}) for the same
matrix dimensions.
Here the agreement is already quite good even for relatively
small matrix sizes. In both plots the errorbars are of order
$10^{-3}$, smaller then the symbols used and therefore omitted.

Concerning the decay of $g_1$, it appears to be
superexponentially in sharp contrast to the algebraic decay
(\ref{eq:CovAsymptAiry2}) of $g_2$. After reinspecting the $2\times2$
block Fredholm determinant this behavior becomes clear, since one of
the off-diagonal blocks is superexponentially small in $u$ for large
values of $u$, while the others stay of order one, a fact already
noticed by Widom 
\cite{Wid07_privatecomm}.  

\begin{figure}
\begin{center}
\includegraphics[height=8cm]{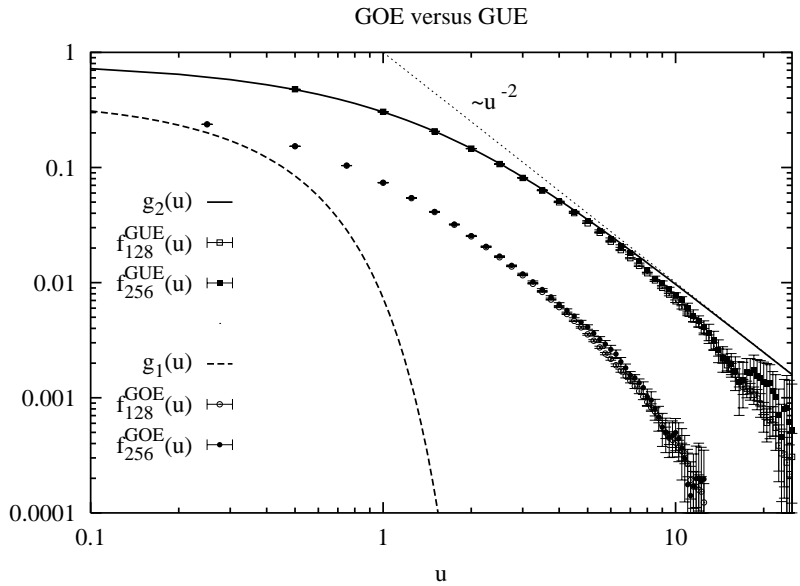}
\caption{Log-log plot of the rescaled correlation functions for GOE and GUE.}
\label{FigCompareGUEGOE}
\end{center}
\end{figure}

Finally, in Figure~\ref{FigCompareGUEGOE}, we provide a comparison of
the decay of correlation for GOE and GUE matrix diffusion. In a
log-log plot we draw $f^{\rm 
  GOE}_N$ and $f^{\rm GUE}_N$ for $N=128$ and $N=256$ with
errorbars. For GUE one observes the deviation from the asymptotic
behavior $u^{-2}$ for large $u$ due to finite size effects.
Remarkably $f^{\rm  GOE}_N(u)$ looks very similar to $\tfrac12f^{\rm
  GUE}_N(2u)$, indicating that the large $u$ behavior
of $f^{\rm GOE}_\infty(u)$ might also be algebraically decaying, in
sharp contrast to the superexponential decay of $g_1(u)$. Given the
small matrix dimensions we used, we can not, however, conclude whether
 the decay
for $f^{\rm GOE}_\infty(u)$ is of order $u^{-2}$ as for GUE or not.

\providecommand{\bysame}{\leavevmode\hbox to3em{\hrulefill}\thinspace}

\end{document}